\documentclass[preprint,showpacs,preprintnumbers,amsmath,amssymb]{revtex4}
\usepackage{graphicx}
\usepackage{dcolumn}
\usepackage{bm}
\begin{document}
\title {Exchange and correlation near the nucleus in density
functional theory}
\author {Zhixin Qian}
\affiliation{Department of Physics,
Peking University, Beijing 100871, China}
\date{\today}
\begin{abstract}
The near nucleus behavior of the exchange-correlation potential
$v_{xc}({\bf r})$ in Hohenberg-Kohn-Sham density functional theory
is investigated.
It is shown that near the nucleus
the linear term of $O(r)$ of 
the spherically averaged exchange-correlation potential 
${\bar v}_{xc}(r)$ is nonzero, and that 
it arises purely from the difference between the
kinetic energy density at the nucleus 
of the interacting
system and the noninteracting Kohn-Sham
system. An analytical expression for the linear
term is derived. Similar results for
the exchange $v_{x}({\bf r})$ and correlation
$v_{c}({\bf r})$ potentials are also obtained
separately. It is further pointed out
that the linear term in $v_{xc}({\bf r})$ arising mainly from
$v_{c}({\bf r})$ is rather small, and $v_{xc}({\bf r})$
therefore has a nearly quadratic structure near the nucleus.
Implications of the results for the construction of
the Kohn-Sham system 
are discussed with examples.
\end{abstract}
\maketitle

Hohenberg-Kohn-Sham density functional theory (HKS-DFT)
\cite{HKS} is now
possibly the most widely used tool 
for determining the electronic structure
of atoms, molecules, and solids \cite{dreizler,parr}.
In the implementations of HKS-DFT,
the so-called Kohn-Sham (KS) noninteracting system which has
electron density $\rho({\bf r})$
equivalent to that of the interacting system is assumed.
For the ground state of an (interacting) $N$-electron system
in an external potential $v_{ext}({\bf r})$, the one-particle
orbitals $\phi_n({\bf r})$ ($n=1, \dots N$) of the corresponding KS system
satisfy the KS equation
\begin{eqnarray} \label{KSequation}
[-\bigtriangledown^2/2 +v_{ext}({\bf r})+ v_H({\bf r}) 
+v_{xc}({\bf r})] \phi_n({\bf r})
= \epsilon_n \phi_n({\bf r}) ,
\end{eqnarray}
with $\epsilon_n$ the eigenvalue.
In Eq. (\ref{KSequation}), $v_H({\bf r})$ is
the classical Hartree potential,
$v_H({\bf r}) = \int d {\bf r}' \rho({\bf r}')/|{\bf r} - {\bf r}'|$,
and $v_{xc} ({\bf r})$ is the KS exchange-correlation ($xc$)
potential 
which is
the derivative of the KS exchange-correlation energy functional
$E_{xc}^{KS}[\rho]$ with respect to the density $\rho({\bf r})$, 
$v_{xc}({\bf r})= 
\delta E_{xc}^{KS}[\rho]/\delta \rho({\bf r})$ \cite{HKS,dreizler,parr}.
The KS $xc$ potential takes account of all the correlations in HKS-DFT.
These correlations arise from the Pauli exclusion principle, Coulomb repulsion, 
and correlation-kinetic effects.
The former two effects are usually characterized with the 
Fermi-Coulomb hole \cite{FChole}, and the latter arises from the
difference between the
kinetic energy density 
of the interacting
and the KS systems. 
$E_{xc}^{KS}[\rho]$
is usually separated into two parts, the exchange $E_{x}^{KS}[\rho]$
and correlation
$E_{c}^{KS}[\rho]$ energy functionals. Correspondingly,
$v_{xc}({\bf r})$ is split into two components,
the KS exchange
$v_{x}({\bf r})$ and correlation $v_{c}({\bf r})$
potentials \cite{HKS,dreizler,parr}.

In practice, an approximation has to be made for the unknown
$E_{xc}^{KS}[\rho]$. In this respect
knowledge of the exact properties of $v_{xc} ({\bf r})$ 
is rather valuable
to the
construction of the KS system.
In this paper it is shown that near nucleus,
\begin{eqnarray}\label{vxc(r)}
{\bar v}_{xc}(r)= {\bar v}_{xc}(0) +
\frac{4}{3}Z[{\bar t}(0) -{\bar t}_s(0)] r/\rho(0) + O(r^2).
\end{eqnarray}
We use ${\bar f}(r)$ to denote the spherical average of a function
$f({\bf r})$.
Here $Z$ is the charge of the nucleus, and $t({\bf r})$ and $t_s({\bf r})$ 
are the kinetic energy
densities for the interacting system
and the KS system, respectively. It is further shown that
\begin{eqnarray}\label{vx}
{\bar v}_{x,c}(r)= {\bar v}_{x,c}(0) +
\frac{4}{3}Z{\bar t}_{x,c}(0)
r/\rho(0) + O(r^2).
\end{eqnarray}
(By the notation ${\bar v}_{x,c}(r)$ we mean ${\bar v}_x (r)$ and
${\bar v}_c (r)$ respectively.)
Here $t_x({\bf r})$ is the first order correction
to $t_s({\bf r})$ in the adiabatic coupling constant perturbation
scheme, (see later discussion and Ref. \cite{harris},)
and $t_c({\bf r})=  t_{xc}({\bf r})- t_x({\bf r})$. Here
$t_{xc}({\bf r})$ is defined to be $t({\bf r}) -t_s({\bf r})$.
The result in Eqs. (\ref{vxc(r)}) and (\ref{vx})
is valid in general whether
the system is an atom, molecule or a solid.

It is well known that,
in the classically forbidden region of finite systems, 
$v_{xc}({\bf r})$ decays asymptotically as $-1/r$ \cite{almbladh},
a result which has played an important role in the
evolution of the energy functional construction.
In contrast, little is known about
$v_{xc}({\bf r})$ near the nucleus.
Inaccuracies in numerical calculations or unreasonable approximations
made could even lead to divergent results at the nucleus,
(see later discussion and Ref. \cite{diverg}.)
It was recently proved that $v_{xc}({\bf r})$
is finite at the nucleus.\cite{vxc(0)}
Whether it behaves linearly or quadratically near the nucleus is
not clear \cite{finite,umrigar}.
In this paper it is pointed out that $v_{xc}({\bf r})$ 
approaches the nucleus
linearly, as shown in Eq. (\ref{vxc(r)}).

The asymptotic $-1/r$ structure in the classically forbidden region
is totally attributable to the
Pauli correlation and
it has been illustrated to arise from the Fermi hole \cite{QS1}. 
The Coulomb correlation
only makes a contribution to the next order 
of $O(1/r^4)$ arising from the Coulomb hole \cite{QS1}. 
It has been 
shown that the correlation-kinetic effects do make a contribution 
asympotically, but only to the order of $O(1/r^5)$ \cite{QS1}. 
Hence in that region
the correlation-kinetic effects
are negligibly small. By contrast Eq. (\ref{vxc(r)})
indicates that, near the nucleus, the linear term of the spherically
averaged $xc$ potential ${\bar v}_{xc}(r)$ arises entirely 
from the correlation-kinetic
effects. (This understanding can also be reached
via Quantal density functional theory (Q-DFT) \cite{qiansahni}.
In Ref. \cite{qiansahni} it is explicitly shown that the
Fermi-Coulomb hole makes no contribution to the linear term
of ${\bar v}_{xc}(r)$.) 
We shall further argue that
the linear term of $v_{xc}({\bf r})$ is rather small,
and $v_{xc}({\bf r})$ is therefore nearly quadratic near the nucleus.

The understanding reached in this work should be 
useful in the construction of the KS system, especially
via approximate $xc$ energy functionals. We shall
demonstrate this with examples after the main derivations.

We start with deriving the following cusp relation for
the density and the kinetic energy density at the nucleus
for the interacting system,
\begin{eqnarray}\label{cusp} 
5Z{\bar \rho}''(0) + {\bar \rho}'''(0) = \frac{8}{3} Z [4Z^2 \rho(0) 
+ {\bar t}(0)],
\end{eqnarray}
and the corresponding one for the KS system,
\begin{eqnarray} \label{KScusp}
5Z{\bar \rho}''(0) + {\bar \rho}'''(0) =&& \frac{8}{3}Z 
[4Z^2\rho(0) + {\bar t}_s(0)     \nonumber \\
&&+ \frac{3}{4Z}\rho(0)({\bar v}'_{H}(0)+ {\bar v}'_{xc}(0))] ,
\end{eqnarray}
where the double and triple primes denote
the second and third derivatives, respectively, with respect to $r$. 
To this end, 
we expand the ground state many-body wavefunction for the 
interacting $N$ electron
system as
\begin{eqnarray}\label{wavefunction}
\Psi({\bf r}, {\bf X})&& =  \Psi(0, {\bf X}) + a({\bf X})r
+ b({\bf X})r^2 + c({\bf X}) r^3 + \dots   \nonumber \\
&& + \sum_{m=-1}^1[a_{1m}({\bf X})r 
+ b_{1m}({\bf X})r^2] Y_{1m}({\hat r}) + \dots  \nonumber \\
&& + \sum_{m=-2}^2 b_{2m}({\bf X})r^2 Y_{2m}({\hat r}) + \dots,
\end{eqnarray}
for small $r$, where ${\hat r}={\bf r}/r$, and ${\bf X}$
denotes $s, {\bf r}_2 s_2, \dots, {\bf r}_N s_N$. 
By substituting the wavefunction into
the many-body Schr\"odinger equation, one obtains the following 
relations:
\begin{eqnarray}\label{a-Psi(0)}
&&a({\bf X}) + Z \Psi(0, {\bf X})=0 ,  \nonumber \\
&&2 b_{1m}({\bf X}) + Za_{1m}({\bf X}) =0 , \nonumber \\
&&4Z b({\bf X}) -Z^3 \Psi(0, {\bf X}) +6 c({\bf X}) =0 .
\end{eqnarray}
These relations originate from the electron-nucleus cusp
of the wavefunction, and 
were derived previously in Ref. \cite{kato}. 
With them, one can see that the density behaves as
\begin{eqnarray} \label{rho1}
{\bar \rho}(r) &=& [(1-Zr)^2 +Z^3r^3/3] \rho(0)   \nonumber \\
&&+ 2(r^2 -5Zr^3/3 )N \int d{\bf X}
Re[\Psi^*(0, {\bf X}) b({\bf X})]  \nonumber \\
&&+(r^2 -Zr^3) N \int d{\bf X} 
\sum_{m=-1}^{1}|a_{1m} ({\bf X})|^2/4 \pi \nonumber \\
&&+ O(r^4) .
\end{eqnarray}
The kinetic energy density,
$t({\bf r}) =
\frac{1}{2} N \int d {\bf X} \bigtriangledown \Psi^*({\bf r}, {\bf X})
\cdot \bigtriangledown \Psi({\bf r}, {\bf X})$,
behaves as
\begin{eqnarray} \label{t(r)}
{\bar t}(r) =& & {\bar t}(0) -\frac{2}{3}Zr [2{\bar t}(0)
-Z^2 \rho(0)]   \nonumber \\
& & -2Zr N  \int d {\bf X} Re
[\Psi^*(0, {\bf X}) b({\bf X})] + O(r^2),
\end{eqnarray}
where 
\begin{eqnarray} \label{t(0)}
{\bar t}(0) = \frac{1}{2} Z^2 \rho(0) + N \int d {\bf X} \sum_{m=-1}^{1}
\frac{3}{8 \pi} |a_{1m} ({\bf X})|^2 .  
\end{eqnarray}
We note that the second term on the right hand side ({\em rhs}) of 
the preceding expression for ${\bar t}(0)$
is absent in the literature \cite{dreizler}.
Combining Eqs. (\ref{rho1}) and (\ref{t(0)}) yields
the relation of Eq. (\ref{cusp}).

Next we consider the corresponding KS system. 
We write $\phi_n({\bf r})$ 
as
\begin{eqnarray} \label{KSorbital}
\phi_n({\bf r})= \sum_{lm} 
r^l [&& A_{nlm} +B_{nlm} r + C_{nlm} r^2   \nonumber \\
&&+ D_{nlm} r^3 + \dots ] Y_{lm} ({\hat r}) .
\end{eqnarray}
Substituting Eq. (\ref{KSorbital}) into Eq. (\ref{KSequation})
with $v_{ext}({\bf r}) = -Z/r$,
one can obtain
\begin{eqnarray} \label{B00-A00}
&&B_{n00} + Z A_{n00} =0 ,  \nonumber \\
&&2B_{n1m} + Z A_{n1m} =0 , \nonumber \\
&&4ZC_{n00} - (Z^3 + {\bar v}_{H}'(0)
+{\bar v}_{xc}'(0) ) A_{n00} + 6 D_{n00} =0 . \nonumber \\
\end{eqnarray}
Evidently these relations similarly 
originate from the electron-nucleus cusp of the KS orbitals. 
Substituting Eq. (\ref{KSorbital}) into the expression for
the density for the KS system, 
$\rho({\bf r}) = \sum_{n=1}^N |\phi_n({\bf r})|^2$, together with
the relations in Eq. (\ref{B00-A00}), one has
\begin{eqnarray} \label{rho4}
{\bar \rho} (r) &&= \rho(0) [(1-Zr)^2 +
\frac{1}{3} (Z^3 +{\bar v}_{H}'(0)
+{\bar v}_{xc}'(0)) r^3 ]  \nonumber \\
&&+\frac{1}{2 \pi} \sum_{n=1}^N Re (A_{n00}^*C_{n00})
(r^2 -\frac{5}{3}Zr^3)     \nonumber \\
&&+\frac{1}{4 \pi} (r^2 -Zr^3) \sum_{n=1}^N
\sum_{m=-1}^1 |A_{n1m}|^2 +O(r^4) .
\end{eqnarray}
Similarly, the kinetic energy density for the KS system,
$t_s({\bf r})= \frac{1}{2} 
\sum_{n=1}^N \bigtriangledown \phi_n^*({\bf r})
\cdot \bigtriangledown \phi_n({\bf r})$,
behaves as
\begin{eqnarray} \label{ts(r)}
{\bar t}_s(r) =& &{\bar t}_s(0) -\frac{2}{3}Zr [2{\bar t}_s(0)
-Z^2 \rho(0)]   \nonumber \\
   \nonumber \\
& &- 2Zr \sum_{n=1}^N \frac{1}{4 \pi} Re [A_{n00}^*C_{n00}] 
+ O(r^2),
\end{eqnarray}
where 
\begin{eqnarray} \label{ts(0)}
{\bar t}_s (0) = \frac{1}{2} Z^2 \rho(0) + \frac{3}{8 \pi}
\sum_{n=1}^N \sum_{m=-1}^{1}|A_{n1m}|^2  .
\end{eqnarray}
Combining Eqs. (\ref{rho4}) and (\ref{ts(0)})
leads to Eq. (\ref{KScusp}).
Once again, we note that
the second term on the {\em rhs} of Eq. (\ref{ts(0)}) for
${\bar t}_s (0)$ is absent in the 
literature \cite{dreizler}. It is exactly
this term and the corresponding term in ${\bar t} (0)$
on the {\em rhs} of Eq. (\ref{t(0)}) that make ${\bar t} (0)
\neq {\bar t}_s (0)$, a fact which is critical to the nonzero linear
term of ${\bar v}_{xc}(r)$. In fact,
a comparison of Eq. (\ref{cusp}) and Eq. (\ref{KScusp}) leads to
${\bar v}'_{H}(0)+{\bar v}'_{xc}(0)
= 4Z[ {\bar t}(0) - {\bar t}_s(0)] /3\rho(0)$,
which equivalently implies Eq. (\ref{vxc(r)}), since ${\bar v}'_H(0)=0$.

Only the properties of ${\bar t}(0)$ and
${\bar t}_s(0)$ have been employed in the preceding derivation,
while accuracy to the linear terms in ${\bar t}(r)$ in
Eq. (\ref{t(r)}) and ${\bar t}_s(r)$ in Eq. (\ref{ts(r)})
enables us to obtain an interesting
cusp relation,
\begin{eqnarray} \label{txccusp}
{\bar t}'_{xc}(0) = -\frac{2}{3} Z {\bar t}_{xc}(0),
\end{eqnarray}
with the aid of a comparision of
the respective coefficients for the $r^2$ and $r^3$ terms  
in Eqs. (\ref{rho1})
and (\ref{rho4}).

To investigate further the near nucleus 
behavior of the KS exchange $v_x({\bf r})$
and correlation $v_c({\bf r})$ potentials separately, 
we employ the adiabatic
coupling constant perturbation scheme \cite{harris}. 
In such a scheme,
one has, instead of Eq. (\ref{vxc(r)}),
\begin{eqnarray}\label{vee(r)-lambda}
{\bar v}_{xc}^{\lambda}(r)= {\bar v}_{xc}^{\lambda}(0) +
\frac{4}{3}Z[{\bar t}^{\lambda}(0) 
-{\bar t}_s(0)] r/\rho(0) + O(r^2).
\end{eqnarray}  
On the other hand, it has been shown that \cite{gorlinglevy}
\begin{eqnarray}  \label{vee-lambda}
v_{xc}^{\lambda}({\bf r})= \lambda v_x({\bf r}) + v_c^\lambda ({\bf r}) ,
\end{eqnarray}
and $v_c^\lambda ({\bf r})$ commences in second order of $\lambda$. 
By comparing Eqs. (\ref{vee(r)-lambda}) and (\ref{vee-lambda}),
one arrives at Eq. (\ref{vx}).
In the meanwhile one has
\begin{eqnarray} \label{tx-ccusp}
{\bar t}'_{x,c}(0) = -\frac{2}{3} Z {\bar t}_{x,c}(0),
\end{eqnarray}
in addition to Eq. (\ref{txccusp}).

It is well known that $t_x({\bf r})$ integrates to zero, i.e., $T_x = 
\int d {\bf r} t_x({\bf r})=0$. Notice that $t_x({\bf r})$ is however not
necessarily zero except for
homogeneous systems. Nevertheless, the fact that $T_x =0$
somewhat indicates that the linear term of ${\bar v}_{xc} (r)$
mainly arises from ${\bar v}_{c}(r)$, according to
Eq. (\ref{vx}). In other words, the linear term
is rather small, and $v_{xc} ({\bf r})$
is nearly quadratic near the nucleus. This argument
might be corroborated by the fact
that both the following sphericalized
approximate exchange potentials approach the
nucleus quadratically: (a) the Pauli correlated
approximation $W_x({\bf r})$ of Q-DFT \cite{harbola}, which is the part of
the exchange potential arising purely from the Fermi hole,
\begin{eqnarray} \label{Wx}
{\bar W}_x (r) = {\bar W}_x (0) + O(r^2) ;
\end{eqnarray}
(b) the Krieger-Li-Iafrate (KLI) approximation
\cite{KLI} to the optimized exchange potential (OEP) \cite{OEP},
\begin{eqnarray} \label{KLI}
{\bar v}_x^{KLI} (r) = {\bar v}_x^{KLI} (0) + O(r^2) .
\end{eqnarray}
Equation (\ref{Wx}) follows directly from the fact that the Fermi hole
for an electron at the nucleus of a sphericalized system
is spherically symmetric \cite{qiansahni}.
Equation 
(\ref{KLI}) can be analytically derived.
(It is worth mentioning that the exact OEP, being the 
exact exchange-only scheme \cite{OEP}, 
approaches the nucleus linearly according
to the result of this paper.)
Therefore, in contrast to the common wisdom that
$v_x({\bf r})$ dominates $v_c({\bf r})$ as is the case in most
other regions, 
$v_{c}({\bf r})$ plays a much more significant
role in the linear term of $v_{xc}({\bf r})$ near the nucleus.

A nearly exact result for the density of the helium atom
is available, which makes it an excellent testing ground for
various approximate energy functionals.
The single
occupied KS orbital for the two electrons with opposite spins is simply
$\phi({\bf r}) = \sqrt{\rho({\bf r})/2}$.
The kinetic energy density at the nucleus of the KS system
can be readily shown, by employing this orbital, to be
\begin{eqnarray} \label{he-ts}
t_s(0) = \frac{1}{2}Z^2 \rho(0) .
\end{eqnarray}
Therefore, according to Eq. (\ref{vxc(r)}), one has
$v_{xc}'(0)= 2Z[2t(0)/\rho(0) -Z^2]/3$. Further 
employment of the cusp relation
of Eq. (\ref{cusp}) yields
\begin{eqnarray}
v_{xc}'(0)=[5Z{\rho}''(0) + {\rho}'''(0) -12Z^3 \rho(0)]/2\rho(0),
\end{eqnarray}
a result obtained previously in Ref. \cite{umrigar}.
Furthermore, in this case,
$v_x(r) = - v_H(r)/2$. Therefore $v_x(r)$ has no contribution
to the linear order,
and $v_c(r)$ makes the entire contribution to the linear term
of $v_{xc}(r)$.
Notice that interestingly the second 
term on the {\em rhs} of Eq. (\ref{ts(0)})
for $t_s(0)$
is absent in Eq. (\ref{he-ts}) in this special case, since both
electrons occupy the $1s$ orbital. This is in fact also
true for three- or four-electron systems like Li and Be atoms and
their isoelectronic sequences since all electrons occupy
single particle states with $l=0$.  The corresponding
second term on the {\em rhs} of Eq. (\ref{t(0)}) for $t(0)$ is 
however nonzero, which
makes the entire contribution to ${v}'_{xc}(0)$. 

\begin{center}
{ Table 1}\\
{Comparison of the exact results for $v_{x,c}(0)$ and
$v'_{x,c}(0)$ and those calculated with LDAs for
the helium atom. (a), (b) and (c) respectively refer to
Vosko-Wilk-Nusair, Perdew-Wang,
and Wigner parametrizations \cite{LDA} for
the correlation energy functionals. Atomic
units are used.}  \\
\begin{tabular}{ c c c c c }
\hline \hline
\\
\ \  ~~~~ \ \  &  \ \ \  $ v_x(0) $ \ \ \ \  & \ \ \ \ $ v_c(0)$ \ \ \ \ & \
 \ \ \ $v'_x(0)$ \ \ \ \ &  \ \ \ \ $v'_c(0)$ \ \ \ \
\\
\hline
Exact &    -1.69  &   -0.062         &    0     & 3.8 \\
\hline
~~ &  ~~  &   -0.092 (a)      &    ~~     &   0.036 (a)  \\
LDA  &  -1.51    &   -0.082 (b)       &    2.02       &   0.035 (b)   \\
~~  &   ~~   &   -0.055 (c)   &    ~~       &   0.0025 (c) \\
\hline
\end{tabular}
\end{center}

All the commonly used generalized 
gradient approximation (GGA) functionals
suffer from spurious divergence 
for $v_{xc}({\bf r})$ at the nucleus \cite{diverg,umrigar}.
This is also true for the later proposed GGA in Ref. \cite{pbe}.
The divergence originates from terms in
$v_{xc}({\bf r})$ containing factor $\bigtriangledown^2 \rho({\bf r})$
which has behavior of $O(1/r)$ at the nucleus.
The local density approximation (LDA), on the other hand,
yields reasonable result for
$v_x(0)$ and $v_c(0)$ as shown in Table 1 for
the helium atom. But, as shown
in Table 1, the LDA for the exchange
component largely overestimates $v'_x(0)$, and the three
LDAs proposed for the correlation component
all hugely underestimate $v'_c(0)$, against
the fact revealed in this paper that the correlation
component plays a dominant role in
the linear term of $v_{xc}({\bf r})$ near the nucleus.

Finally we take the electrostatic interpretation of the KS $xc$ potential
\cite{qxc} as
another example to illustrate the possible implications of our investigation
for the
construction of the KS system. In this interpretation, the concept of
static exchange-correlation charge density $q_{xc}({\bf r})$ is introduced
for the KS $xc$ potential,
\begin{eqnarray}
\bigtriangledown^2 v_{xc}({\bf r}) = - 4 \pi q_{xc}({\bf r}).
\end{eqnarray}
The result shown in Eq. (\ref{vxc(r)}) indicates that $q_{xc}({\bf r})$
diverges at the nucleus,
\begin{eqnarray}
{\bar q}_{xc}(r \to 0) 
= -\frac{2}{3 \pi r}Z[{\bar t}(0) -{\bar t}_s(0)] /\rho(0).
\end{eqnarray}
This unphysical feature reveals inherent shortcomings  of the electrostatic
interpretation of $v_{xc}({\bf r})$. It 
implies that the part of $v_{xc}({\bf r})$
due to the correlation-kinetic effects can not be properly
interpreted in terms of a static charge density. 
Furthermore, according to the 
result in Eq. (\ref{vx}), the shortcomings persist
in the analogous 
interpretation \cite{qxc} of the exchange $v_x({\bf r})$ and 
correlation $v_c({\bf r})$ 
potentials, separately, in terms of static exchange $q_{x}({\bf r})$
and correlation $q_{c}({\bf r})$ charge densities.
 
It might be necessary to point out the following fact before
we close the paper. As mentioned in the introduction,
in addition to the traditional exchange-correlation energy
$E_{xc}[\rho]$ of the conventional many-body theory, which arises
purely from the interaction between an electron and 
its Fermi-Colomb hole, there is a contribution of the 
correlation-kinetic energy $T_{xc}[\rho]=\int d{\bf r}
[t({\bf r}) - t_s({\bf r})]$ to
the KS exchange-correlation energy: $E_{xc}^{KS}[\rho] 
=E_{xc}[\rho] + T_{xc}[\rho]$. Correspondingly, $v_{xc} ({\bf r}) =
\delta E_{xc}[\rho]/\delta \rho + \delta T_{xc}[\rho]/\delta \rho$. 
The linear term in ${\bar v}_{xc}(r)$ in Eq.
(\ref{vxc(r)}) is identified 
as the correlation-kinetic effects, but
it remains not clear whether it arises entirely from
$\delta T_{xc}[\rho]/\delta \rho$. Nevertheless, the result in this work
indicates that, near the nucleus, $T_{xc}[\rho]$ plays 
a significant role in
the performance of the approximate KS $xc$ energy functionals. 

The author is grateful to Prof. V. Sahni for discussions and help,
and Prof. C. J. Umrigar for providing accurate data on the
density of the helium atom.
This work was supported by the Chinese National Science Foundation
under Grant No. 10474001.


\begin{thebibliography}{100}
\bibitem{FChole}
Here we mean the traditional Fermi-Coulomb hole in many-body
theory,
in contrast to the exchange-correlation hole extensively
used in HKS-DFT. The exchange-correlation hole is the coupling constant
average of the Fermi-Coulomb hole in the adiabatic coupling
constant scheme, (see Ref. \cite{harris}
and the later discussion in the present paper.)
\end{thebibliography}
\end{document}